\documentclass[notitlepage,amsmath,showpacs,twocolumn,amssymb,aps,pre,10pt,a4paper]{revtex4-1}
\DeclareMathOperator{\tr}{tr}

\DeclareMathOperator*{\Res}{Res}

\providecommand{\mean}[1]{\langle #1 \rangle}
\providecommand{\bra}[1]{\langle #1 |}

\providecommand{\ket}[1]{| #1 \rangle}
\providecommand{\irr}[0]{\text{irr}}

\providecommand{\abs}[1]{\bigl| #1 \bigr|}
\providecommand{\sabs}[1]{\left| #1 \right|}

\usepackage{graphicx,color,bm}
\usepackage[caption=false]{subfig}

\begin{document}

\preprint{SanD-0114}

\title{Exact work statistics of quantum quenches in the anisotropic XY model}% Force line breaks with \\

\author{Francis A.~Bayocboc,~Jr.}
 \email{Current address: Institute of Mathematical Sciences and Physics, University of the Philippines Los Ba\~nos, 4031 Laguna, Philippines}
\author{Francis N.~C.~Paraan}
 \email{fparaan@nip.upd.edu.ph (Corresponding author)}
\affiliation{National Institute of Physics, University of the Philippines Diliman, 1101 Quezon City, Philippines}

\date{\today}% It is always \today, today,

\begin{abstract}
\noindent We derive exact analytic expressions for the average work done and work fluctuations in instantaneous quenches of the ground and thermal states of a one-dimensional anisotropic XY model. The average work and a quantum fluctuation relation is used to determine the amount of irreversible entropy produced during the quench, eventually revealing how the closing of the excitation gap leads to increased dissipated work. The work fluctuation is calculated and shown to exhibit non-analytic behavior as the pre-quench anisotropy parameter and transverse field are tuned across quantum critical points. Exact compact formulas for the average work and work fluctuation in ground state quenches of the transverse field Ising model allow us to calculate the first singular field derivative at the critical field values.
%\begin{description}
%\end{description}
\end{abstract}

\pacs{05.70.Ln, 05.30.-d, 75.10.Pq}

%05.30.-d	Quantum statistical mechanics
%75.10.Pq	Spin chain models
%05.70.Ln	Nonequilibrium and irreversible thermodynamics

\maketitle

\section{Introduction}

A quantum quench (or non-adiabatic change in the system Hamiltonian) of an isolated many-body system generally leads to a state that is far from equilibrium. In particular, a thermal state $\rho_0 = e^{-\beta H_0}/\tr e^{-\beta H_0}$ does not generally evolve into a canonically distributed (Gibbs) state by unitary dynamics under a different Hamiltonian $H_1$ \cite{rigol2008a}. There is much interest in the study of these highly excited quenched states, especially when the quench is done about a quantum critical point, because of the possibility of observing universal phenomena \cite{polkovnikov2011a,jacobson2011a,gambassi2012a,mazza2012a,palmai2014a}, dynamical phase transitions \cite{heyl2013a}, quantum revivals or thermalization \cite{igloi2000a,greiner2002a,sengupta2004a,kinoshita2006a,calabrese2011e,happola2012a,eisert2015a}, and singular behavior in the thermodynamic limit \cite{silva2008a,calabrese2012m,calabrese2012n,dorner2012a,shchadilova2014a,palmai2015a}. Also, when the quench involves interaction parameters that couple different subsystems within a system, non-trivial correlations and entanglement may be generated between these subsystems \cite{fagotti2008a,schachenmayer2013a,fagotti2013a}. Thus, it has been suggested that quench-based protocols may be used to measure entanglement entropies \cite{cardy2011a}.

Furthermore, when a quantum quench is viewed as a thermodynamic process, it can lead to dynamically-generated quantum fluctuations that are described by generalized fluctuation relations \cite{esposito2009a,campisi2011a,hanggi2015a}. For example, the work done on a quenched isolated system is described by a probability distribution whose characteristic function takes the form of a two-time correlation function \cite{talkner2007a}. This correlation function can be expressed in terms of free energy differences of a system in effective thermal equilibrium \cite{jarzynski1997a,crooks1999a}. This emerging thermodynamic description therefore opens up the possibility of the use of thermodynamic (statistical) methods in the description of isolated quantum systems driven away from equilibrium (cf.~\cite{silva2008a,paraan2009a,degrandi2010a,dorner2012a,calabrese2012n,mascarenhas2014a,marcuzzi2014a,eisert2015a}). 

In this paper, we continue previous investigations on the work done in quenched quantum chains. In particular, our emphasis is on the theoretical analysis of the statistics of work done in arbitrary quenches of the ground and thermal states of the XY model following the formalism developed in Ref.~\cite{talkner2007a}. Work statistics have been studied in other quenched systems such as the transverse field Ising model \cite{silva2008a,dora2012a,dorner2012a,heyl2013a}, the XXZ model \cite{deluca2014a,torresherrera2014a} and integrable field theories \cite{palmai2014a,palmai2015a}, Luttinger liquids \cite{dora2012a}, low-dimensional quantum gases \cite{gambassi2012a,sotiriadis2013a, deluca2014a,shchadilova2014a}, critical spin-boson models \cite{paraan2009a}, and optomechanical systems \cite{brunelli2015a}. These theoretical studies have recently been complemented by experimental interferometric measurements of the probability distribution of work done on quenched gases of two-level systems \cite{dorner2013a, mazzola2013a, batalhao2014a}. Such measurements have also been viewed from a quantum information perspective, where the measurement of work done is considered as a generalized quantum measurement (a positive operator valued measure or POVM) \cite{roncaglia2014a}. 

The results of this study can be divided into two main parts. In the first part, we report a calculation of the average work done in an isolated quenched XY chain (Section \ref{averagework}). When the system is initially in a canonically distributed state $\rho_0$, the quench is associated with the production of a so-called irreversible entropy $\Delta S_\irr$, which may be interpreted as a measure of the irreversibility of the quench \cite{jarzynski2008a,vaikuntanathan2009a,deffner2010a,dorner2012a, plastina2014a,apollaro2014a}. Indeed, in an instantaneous quench $\Delta S_{\irr}$ can be expressed mathematically as the relative entropy (Kullback-Leibler divergence) between the probability distributions of work done in the forward quench protocol and the reversed protocol (with initial and final Hamiltonians swapped) \cite{dorner2012a}. Additionally, $\Delta S_{\irr}$ has been recently proposed as an experimentally accessible quantity that may be used to establish the arrow of time in isolated quantum systems \cite{batalhao2015a}. We give an exact solution for $\Delta S_\irr$ for arbitrary sets of pre- and post-quench Hamiltonian parameters and examine sharp increases in irreversible entropy production about quantum criticality. 

Second, we calculate the fluctuation in the work done in an arbitrary ground state quench and obtain exact contour integral representations in the thermodynamic limit (Section \ref{workfluc}). We find that the work fluctuation is generally not an analytic function of the pre-quench (but not post-quench) Hamiltonian parameters along the quantum critical lines. For quenches done on thermally mixed states, a numerical analysis reveals that this non-analytic behavior is weakened in the sense that singular behavior begins to appear in higher-order derivatives. These results suggest that quench protocols may be used to locate quantum critical points in systems with unknown phase diagrams. 

Past investigations of quenches in the one-dimensional XY model have focused on the time evolution (quench dynamics) of correlation functions \cite{blass2012a,barmettler2010a,calabrese2006b,calabrese2007a,fagotti2014a}, the Loschmidt echo \cite{jacobson2011a}, defect production \cite{blass2012a,cherng2006a, mukherjee2007a,divakaran2009a}, and entanglement measures \cite{blass2012a,fagotti2008a,happola2012a,torlai2014a}. Detailed theoretical analyses of the work statistics have been performed for quenches along the transverse field Ising line of the XY model \cite{silva2008a,dorner2012a}, and we therefore emphasize in this paper the effects of anisotropy on the average work and work fluctuations. Still, we are able to report some new exact formulas for the average work \eqref{eq:workising} and work fluctuation \eqref{isingfluc} along the Ising line to complement these previous studies. The overall contribution of our analytic results is therefore the completion of the study of work statistics of quantum quenches in the full parameter space of the XY model.

\section{Model and derivations}

\begin{figure}[t]
\centering
\includegraphics[width=.7\linewidth]{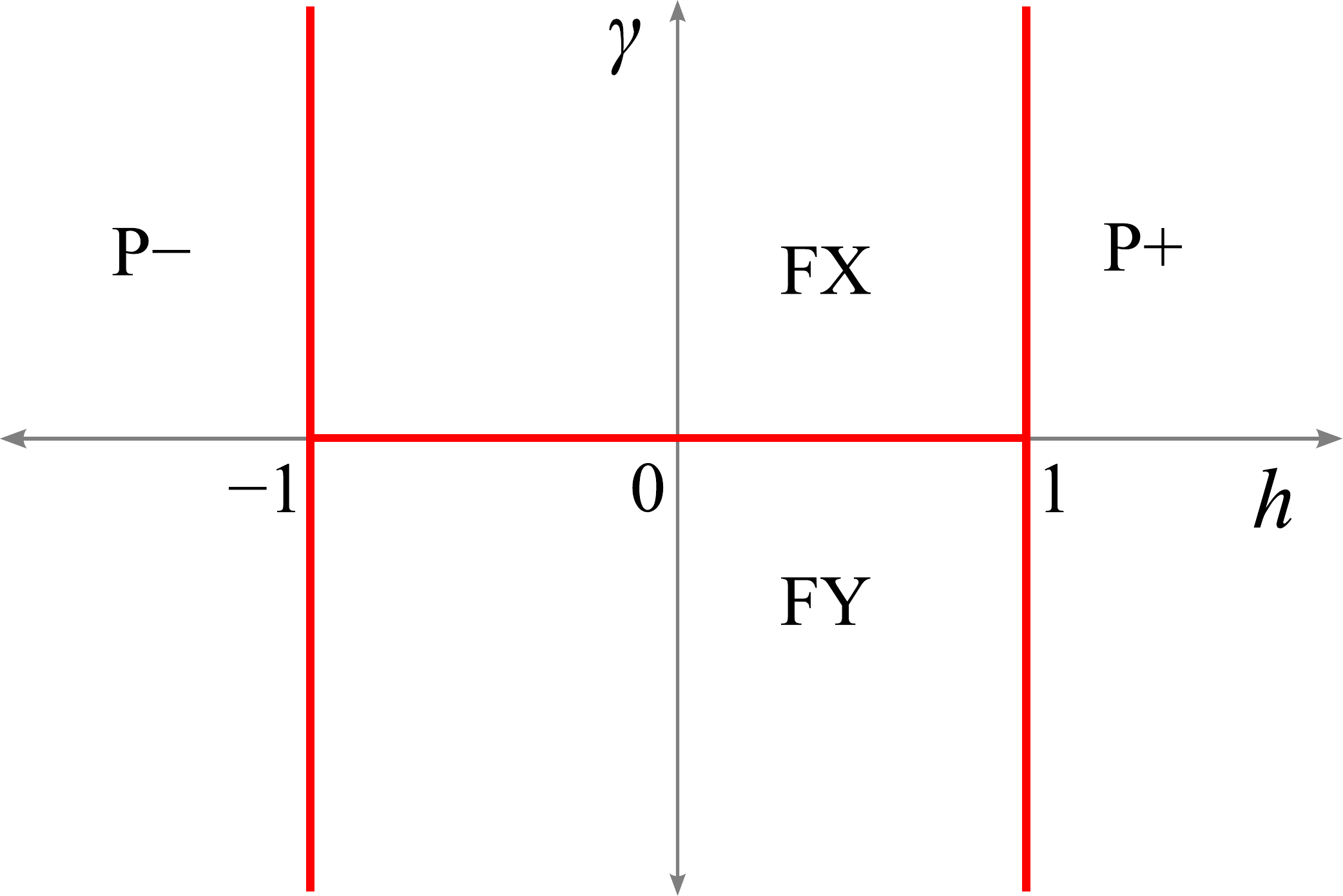}
\caption{(Color online) Phase diagram of the XY model showing the quantum critical lines (red solid lines). The anisotropy parameter is $\gamma$ and the transverse magnetic field is $h$. In the FX and FY phases the ground state has ferromagnetic order in the $x$- and $y$-directions, respectively. The paramagnetic P$\pm$ ground state has no long-range order on the longitudinal $xy$-plane.
}\label{fig:xy_phase_diagram}
\end{figure}

\textit{XY model and quench protocol---}The one-dimensional ferromagnetic XY model Hamiltonian is
\begin{equation}
H = -\frac{1}{2}\sum_{j=1}^N \biggl[\frac{1+\gamma}{2}\sigma_j^x\sigma_{j+1}^x + \frac{1-\gamma}{2}\sigma_j^y\sigma_{j+1}^y + h\sigma_j^z\biggr],
\end{equation}
with anisotropy parameter $\gamma$, transverse field $h$, and $\sigma_j^a$ the Pauli $a$--matrix on site $j$. We impose periodicity on the spin chain so that $\sigma_{N+1}^a = \sigma_1^a$.

The XY Hamiltonian can be diagonalized by a sequence of transforms: a Jordan-Wigner transform to spinless fermionic operators, a Fourier transformation into momentum space, and a Bogolyubov rotation to obtain a non-interacting fermion theory. The final result is 
\begin{equation}\label{Btransformed}
H = \sum_{n=0}^{N-1}\epsilon(q_n) \left(\chi_n^\dagger\chi_n - \tfrac{1}{2} \right),
\end{equation}
where $\chi_n^\dagger$ and $\chi_n$ are canonical fermion creation and annihilation operators, respectively \cite{katsura1962a}. The dispersion relation $\epsilon(q_n)$ is 
\begin{equation}\label{dispersion}
\epsilon(q_n) = \bigl[(h-\cos q_n)^2 + \gamma^2\sin^2 q_n\bigr]^{1/2},
\end{equation}
and $q_n = 2\pi n/N$ for odd $N$ or $q_n= 2\pi (n+\frac{1}{2})/N$ for even $N$. In the thermodynamic limit $N\to\infty$ the effects of $N$ being odd or even are negligible.

The phase diagram of the XY model is given in Figure~$\ref{fig:xy_phase_diagram}$. The excitation gap vanishes along the critical field lines ($h=\pm 1$) and the critical isotropic line ($\gamma = 0$ and $\sabs{h}<1$) and the system is quantum critical in these regions.

We consider here quench protocols in which the initial Hamiltonian $H_0$ has anisotropy parameter $\gamma_0$ and field $h_0$ and is instantaneously changed to the final Hamiltonian $H_1$ with corresponding parameters $\gamma_1$ and $h_1$. This quench is performed with the system initially prepared in a canonically distributed mixed state $\rho_0 = e^{-\beta H_0}/\tr e^{-\beta H_0}$ at  an effective reciprocal temperature $\beta = (kT)^{-1}$, or in the ground state $\rho_0 = \ket{0}\bra{0}$ of $H_0$ at effective zero temperature $\beta\to\infty$. The system is isolated and is not in contact with a thermal bath during the quench. The work done $W$ during the quench is defined as the difference between projective measurements of the system energy after and before the quench.

\textit{Work statistics---}The statistical properties of the probability distribution of work done $p(W)$ in an instantaneous quench are most accessible from the characteristic function
\begin{equation}\label{charequation}
	G(u,\beta) = \int e^{iuW} p(W)\,dW = \tr e^{iuH_1}e^{-iuH_0}\rho_0,
\end{equation}
or for a ground state quench $G(u) = \bra{0}e^{iuH_1}e^{-iuH_0}\ket{0}$ \cite{talkner2007a}. Using the known eigenstates of the XY model, the zero temperature characteristic function $G(u)$ can be calculated and expressed compactly as
\begin{equation}
G(u)= e^{iu\Delta E} \prod_n \bigl[ \cos^2\Delta_n + e^{iu\epsilon_1(q_n)}\sin^2\Delta_n\bigr].
\end{equation}
Here, $\Delta E = E_1-E_0$ is the difference between the post- and pre-quench ground state energies and $\Delta_n = \theta_1(q_n) -\theta_0(q_n)$ is the difference between the post- and pre-quench Bogolyubov angles that satisfy
\begin{equation}
\tan\bigl[2\theta_i(q_n)\bigr] = \frac{\gamma_i\sin q_n}{h_i-\cos q_n}.
\end{equation}
The derivation of this characteristic function follows closely that in the transverse field Ising case where $\gamma_0 = \gamma_1 = 1$ \cite{dorner2012a}. 

The cumulants of the work distribution can therefore be generated by repeatedly differentiating the generating function $\ln G$. In particular, the average work done $\mean{W}$ and variance in work done $\Sigma^2 = \mean{W^2}- \mean{W}^2$ are
\begin{align}
	\mean{W}& = \frac{1}{i}\lim_{u\to0}\frac{\partial}{\partial u} \ln G ,\\
	\Sigma^2&= \frac{1}{i^2}\lim_{u\to0}\frac{\partial^2}{\partial u^2} \ln G.
\end{align}

\section{Average work done and irreversible entropy}\label{averagework}

From the characteristic function \eqref{charequation} we can obtain the average work done in a finite temperature quench:
\begin{equation}\label{eq:averagework}
\langle W \rangle = \frac{1}{2}\sum_{n=0}^{N-1} \bigl[\epsilon_{0}(q_n) - \epsilon_{1}(q_n)\cos 2\Delta_{n} \bigr] \tanh\biggl[ \frac{\beta\epsilon_{0}(q_n)}{2} \biggr].
\end{equation}
In the thermodynamic limit the sum for the average work per spin $\mean{w} \equiv \lim_{N\to\infty}\mean{W}/N$ can be transformed into a definite integral that is solvable by quadratures:
\begin{multline}\label{eq:averageworkthermo}
\langle w \rangle = \frac{1}{4\pi}
\int_0^{2 \pi} \Biggl[\frac{(h_0 - h_1)(h_0 - \cos k) }{\epsilon_0(k)}
\\ \quad + \frac{\gamma_0(\gamma_0 - \gamma_1) \sin^2 k}{\epsilon_0(k)} \Biggr]\tanh\frac{\beta\epsilon_0(k)}{2}\,dk.
\end{multline}
For a ground state quench we take $\beta\to\infty$ before summing or integrating over wave vectors. This exact expression reveals that the average work done can be neatly divided into a contribution from the field change $(h_1 - h_0)$ and a contribution from the anisotropy change $(\gamma_1 - \gamma_0)$. 

\textit{Ground state quench along the Ising line---}For example, in a zero temperature transverse field quench along the Ising line $\gamma_0=\gamma_1 = 1$, the preceding integral at $\beta \to \infty$ can be evaluated exactly in terms of complete elliptic integrals \cite{laurente2015a}:
\begin{equation}\label{eq:workising}
\mean{w} = \frac{h_0-h_1}{\pi h_0} \begin{cases}
	 E({h_0}) - (1-h_0^2) K({h_0}), & \sabs{h_0}<1,\\
	 \sabs{h_0} E({h_0}^{-1}) , & \sabs{h_0}> 1, 
	\end{cases}
\end{equation}
where $K(k)$ and $E(k)$ have the integral representations
\begin{align}
	K(k) &= \int_0^{\pi/2} (1-k^2\sin^2 u)^{-1/2}\,du, \\
	E(k) &= \int_0^{\pi/2} (1-k^2\sin^2 u)^{1/2}\,du.
\end{align}
The exact expression \eqref{eq:workising} for $\mean{w}$ can be cast in the more compact form 
\begin{equation}
	\mean{w} = \frac{h_0-h_1}{\pi h_0} \times \sabs{h_0} E({h_0}^{-1}),
\end{equation}
using the formula $k E(k^{-1}) = E(k) - (1-k^2) K(k)$ \cite{gradshteyn}. Asymptotic series expansions of this result about the pre-quench critical fields $\sabs{h_0}=1$ reveal that the average work per spin is continuous, but not infinitely differentiable, at $\sabs{h_0} = 1$. Specifically, logarithmic singularities are generally observed in the field derivative about the critical field $h_0 = 1$:
\begin{equation}\label{eq:workisingapprox}
\frac{\partial\mean{w}}{\partial h_0} \sim \frac{h_1-1}{2\pi} \times \begin{cases}
	 \ln(1-h_0), & h_0\lesssim 1,\\
	 \ln (h_0-1), & 1 \lesssim h_0 , 
	\end{cases}
\end{equation}
and about the critical field $h_0 = -1$:
\begin{equation}\label{eq:workisingapprox2b}
\frac{\partial\mean{w}}{\partial h_0} \sim \frac{h_1+1}{2\pi} \times \begin{cases}
	 \ln(1-\sabs{h_0}), & \negthickspace -1 \lesssim h_0,\\
	 \ln (\sabs{h_0}-1), & h_0 \lesssim -1.
	\end{cases}
\end{equation}
An exceptional case occurs when the post-quench Hamiltonian is also at critical field. The non-analyticity at $h_0 = h_1 = \pm 1$ is weakened in the sense that the derivative ${\partial\mean{w}}/{\partial h_0}$ exists at critical $h_0$:
\begin{equation}
	\frac{\partial\mean{w}}{\partial h_0}\biggr|_{h_0=\pm 1} = \pm\frac{1}{\pi},\quad \text{for }h_1= \pm 1,
\end{equation}
and now the logarithmic divergence first appears in the second derivative ${\partial^2\mean{w}}/{\partial h_0^2}$. To our knowledge, this is the first exact characterization of the non-analytic behavior of the average work done in a quenched transverse field Ising model at quantum criticality. (We mention, however, that logarithmic singularities in the work fluctuation have been reported in field quenches that are localized to a single spin site \cite{silva2008a}.)

\begin{figure*}[th]
\centering
\quad \subfloat[Field quench $h_1 - h_0 = 0.01$ at constant anisotropy $\gamma$.\label{sfig:fieldquench}]{%
  \includegraphics[width=.45\linewidth]{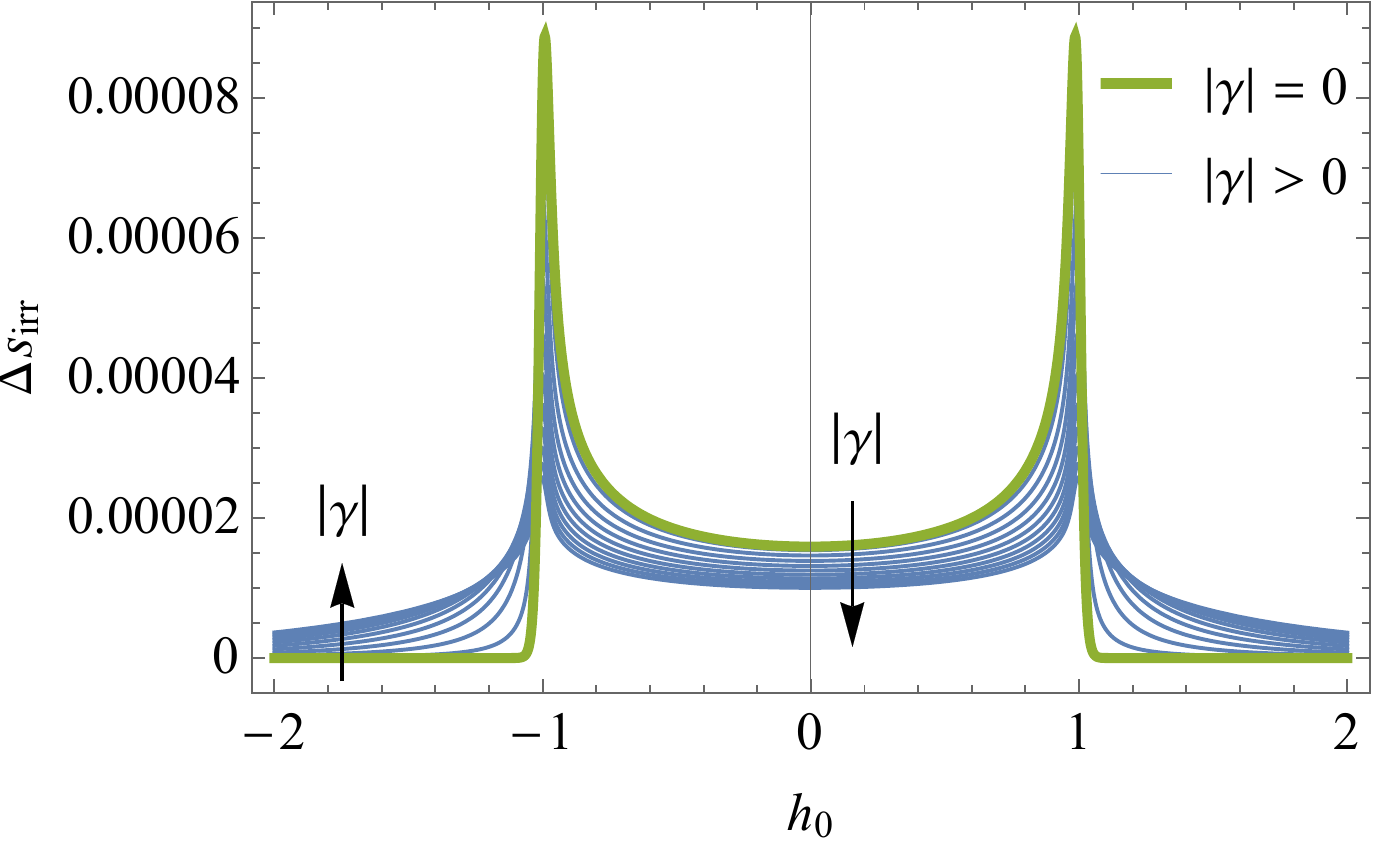}%
}\hfill
\subfloat[Anisotropy quench $\gamma_1 - \gamma_0 = 0.01$ at constant field $h$.\label{sfig:anisotropyquench}]{%
  \includegraphics[width=.45\linewidth]{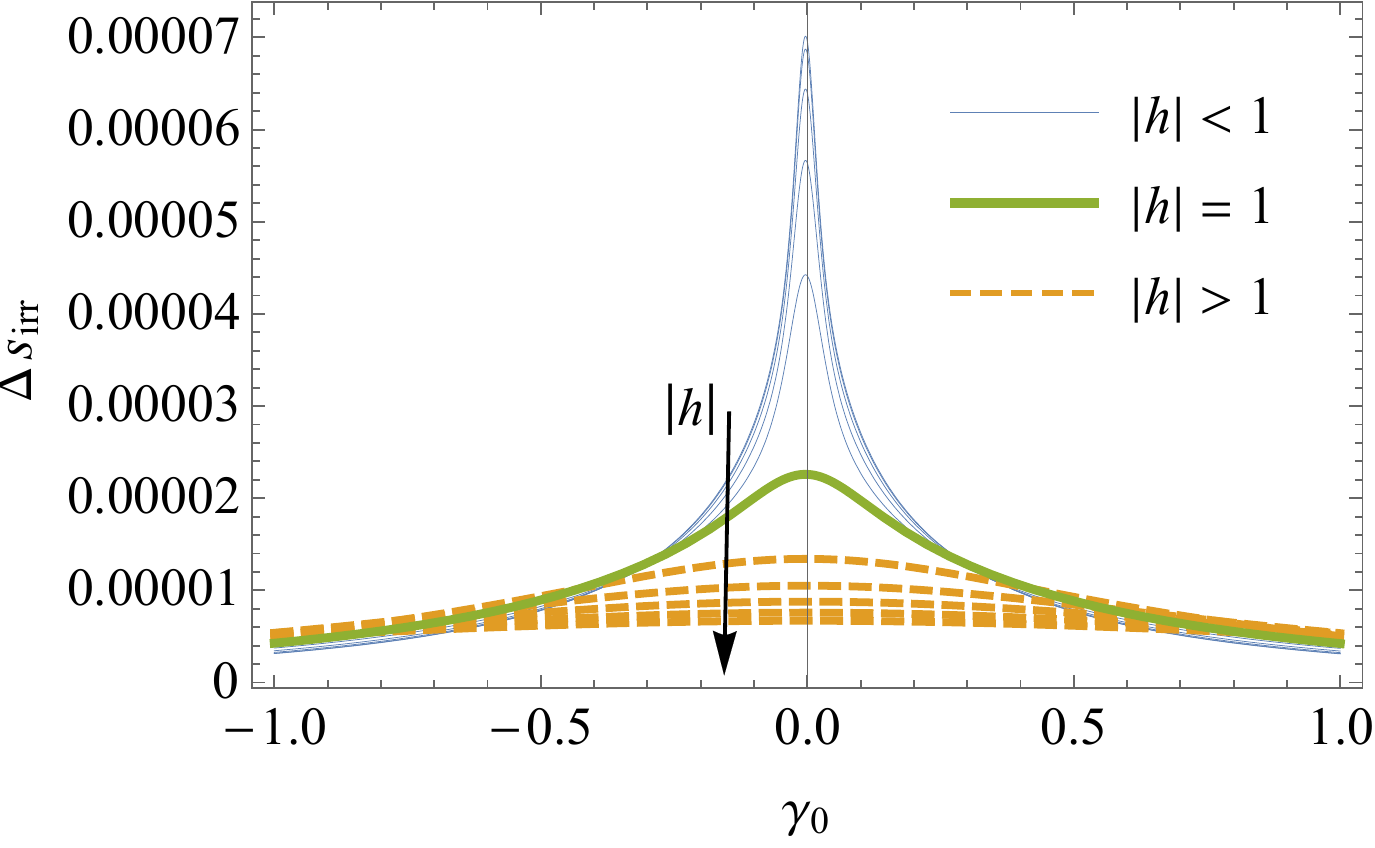}\quad %
}
\caption{(Color online) Irreversible entropy produced per spin $\Delta s_\irr \equiv \lim_{N\to\infty} \Delta S_\irr/N$ at low temperature $\beta=100$ for (a) small field quenches $\delta h = 0.01$, and (b) small anisotropy quenches $\delta\gamma=0.01$. Black arrows on the field (anisotropy) quench graph denote increasing anisotropy $\sabs{\gamma}$ (field $\sabs{h}$) magnitudes from $0$ to $2.0$ in steps of 0.2.}
\end{figure*}

\textit{Irreversible work and entropy---}The average work done in a quantum quench also provides information on the excess work (or irreversible/dissipated work $W_\irr \equiv \langle W\rangle - \Delta F$) done above the free energy change of an analogous reversible process that takes $H_0$ to $H_1$ at reciprocal temperature $\beta$. This free energy change $\Delta F$ is given by the Jarzynski relation \cite{jarzynski1997a}
\begin{equation}
\Delta F = -\frac{1}{\beta}\ln G(i\beta), %= -\frac{1}{\beta}\ln \langle e^{-\beta W} \rangle 
\end{equation}
and may be expressed in terms of effective canonical partition functions ${Z}_i = \tr e^{-\beta H_i}$ through the Tasaki-Crooks fluctuation relation \cite{crooks1999a,talkner2007b}:
\begin{equation}\label{freeenergychange}
{\Delta F} = -\frac{1}{\beta}\ln \frac{Z_1}{Z_0} = -\frac{1}{\beta}\sum_{n=0}^{N-1} \ln \frac{\cosh\bigl[ \beta\epsilon_{1}(q_n)/2 \bigr]}{\cosh\bigl[ \beta\epsilon_{0}(q_n)/2 \bigr]}.
\end{equation}
The irreversible work done is positive as excitations, or defects, are produced during the quantum quench. It is then usually associated with an irreversible entropy production according to $\Delta S_{\irr} = \beta W_\irr= \beta(\mean{W}-\Delta F)$. 

We remark that the expressions derived here for the average work \eqref{eq:averagework} and free energy change \eqref{freeenergychange} are identical in form to the corresponding equations previously obtained for the transverse field Ising model \cite{silva2008a,dorner2012a}. The generalization to the full XY model simply involves the modification of the dispersion relations and Bogolyubov angles to account for the anisotropy $\gamma$. With the full expression however, we are able to numerically investigate the effects of anisotropy on the quenched model and explore critical phenomena along the critical XX line. In the examples that follow, we study how field quenches are modified by anisotropy and how the behavior of anisotropy quenches depend on whether the transverse field is above or below the critical threshold.

First, we consider small transverse field quenches $\delta h = h_1-h_0 = 0.01$ at fixed anisotropy and low temperature $\beta = 100$. Figure~\ref{sfig:fieldquench} shows how the irreversible entropy is pronounced for quenches in the vicinity of the critical field due to the increased amount of irreversible work done in defect production as the energy gap closes \cite{cherng2006a,mukherjee2007a,divakaran2009a}. Accounting for anisotropy in the XY model reveals additional features of $\Delta S_\irr$ than previously observed along the Ising line. For instance, the production of irreversible entropy is greatest about the multicritical points at the intersection of the XX and critical field lines (seen as the peaks at $h_0 = \pm 1$ of the thick solid line in Fig.~\ref{sfig:fieldquench}). Also, we observe the different effect of anisotropy on small $\delta h$ field quenches deep in the ferromagnetic $\sabs{h}<1$ and paramagnetic phases $\sabs{h}>1$. In the former, where ferromagnetic order is established along the $x$- and $y$-directions, increasing anisotropy from $\sabs{\gamma}=0$ decreases the irreversible entropy production, while in the latter increasing anisotropy increases $\Delta s_\irr$. 

Second, we take a look at the case of anisotropy quenches $\delta \gamma = \gamma_1-\gamma_0 = 0.01$ at fixed field and low temperature $\beta = 100$. In Fig.~\ref{sfig:anisotropyquench} we see that irreversible entropy production peaks when the quench is done about the isotropic XX line. However, this peak is most prominent when the quench is done within the critical region $\sabs{h}<1$, again mirroring the increased dissipated work at criticality. However, unlike the previous case of field quenches, less irreversible entropy is produced when the anisotropy quench is done about the multicritical points ($h = \pm 1$) than when done across the critical XX line at low transverse fields $\sabs{h}<1$.  That is, in a given anisotropy quench about the isotropic XX region, an increasing transverse field suppresses the production of excitations.

\begin{figure}[b!]
\centering
\includegraphics[width=0.7\linewidth]{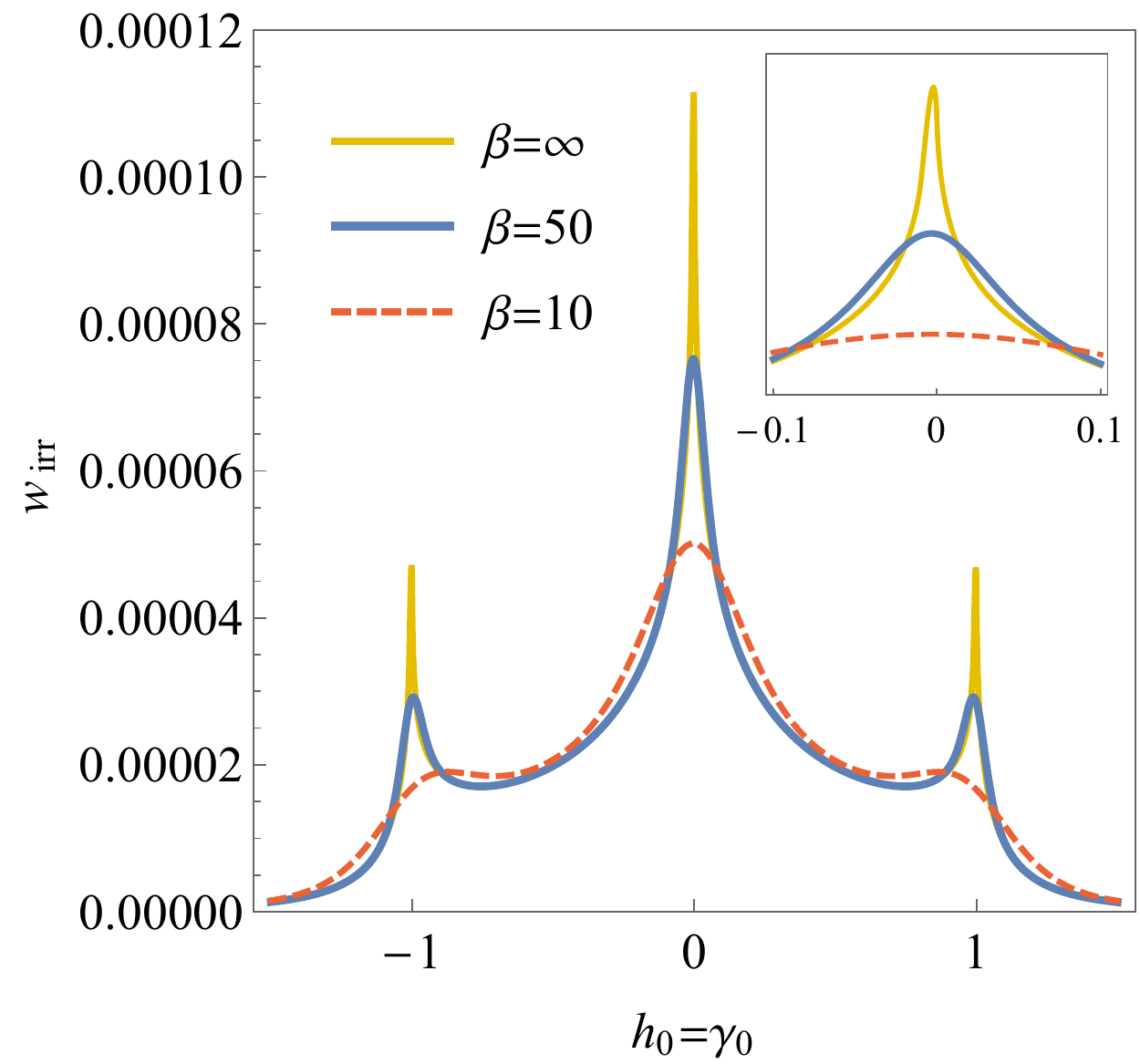}
\caption{(Color online) Irreversible work per spin $w_\text{irr}\equiv \lim_{N\to\infty} W_\text{irr}/N$ for small magnitude ground state quenches $\delta h = \delta \gamma = 0.01$ along the $h_0 = \gamma_0$ line. Three critical lines are crossed, which correspond to peaks in the irreversible work. These peaks become sharper as the temperature is decreased to zero (the inset shows the peak at $h_0 = \gamma_0=0$).}\label{fig:irrentropy}
\end{figure}

\textit{Non-zero temperature---}In the limit of infinite temperature $\beta\to 0$ the irreversible entropy $\Delta S_\irr$ vanishes trivially as $\lim_{\beta\to 0}\beta \mean{W}$ and $\lim_{\beta\to 0}\beta \Delta F$ vanish individually. In this extreme case, the pre- and post-quench states are both maximally mixed and have zero relative entropy. In contrast, when the pre-quench state is prepared at a finite non-zero temperature, the temperature dependence of the average work \eqref{eq:averageworkthermo} and free energy change \eqref{freeenergychange} are difficult to study analytically because of their hyperbolic dependence on the energy spectrum $\epsilon_0$. Still, the integrals for the irreversible entropy and irreversible work per spin are readily evaluated by quadrature methods and we performed empirical studies of the temperature dependence of these quantities. For instance, in the case of small magnitude quenches $\delta h = \delta \gamma = 0.01$ beginning at $h_0 = \gamma_0$, peaks in the irreversible work per spin $w_\irr = \Delta s_\irr/\beta$ can be examined as three critical lines are crossed (Fig.~\ref{fig:irrentropy}). As in the case of quenches along the Ising line \cite{dorner2012a}, these peaks are caused by increased defect production as the excitation gap closes at criticality. Additionally, it turns out that the irreversible work per spin $w_\irr$ saturates to a constant value in the thermodynamic limit at zero temperature, that is, it does not diverge at any temperature. This result implies that the total irreversible entropy, like a thermodynamic entropy, scales extensively at all finite temperatures $\Delta S_\irr\sim N$ as $N\to\infty$. Furthermore, for these small magnitude quenches, the irreversible work vanishes at high fields and large anisotropy at all temperatures as the relative parameter changes $\delta h/h_0$ and $\delta \gamma/\gamma_0$ become smaller \cite{dorner2012a}.

%However, the irreversible entropy per spin $\Delta s_\irr \equiv \lim_{N\to\infty} \Delta S_\irr/N $ can be shown to vanish in the limit $\beta \to 0$. That is, at increasing temperatures the forward and backward distributions of work done become more similar and the quantum quench, when viewed as a thermodynamic transformation, becomes more reversible. 

\begin{figure}[tb]
\centering
 \includegraphics[width=.8\linewidth]{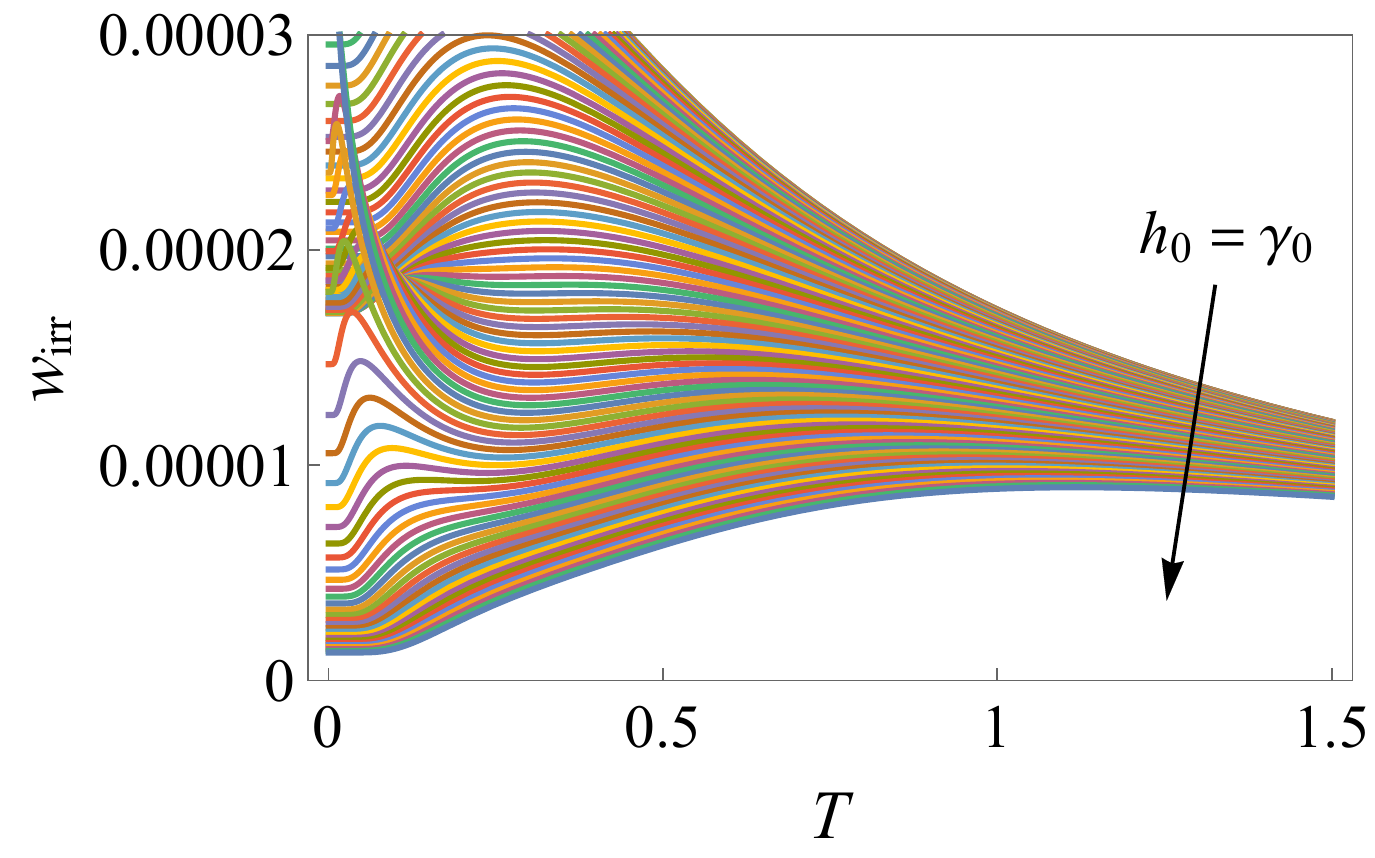}

\includegraphics[width=.7\linewidth]{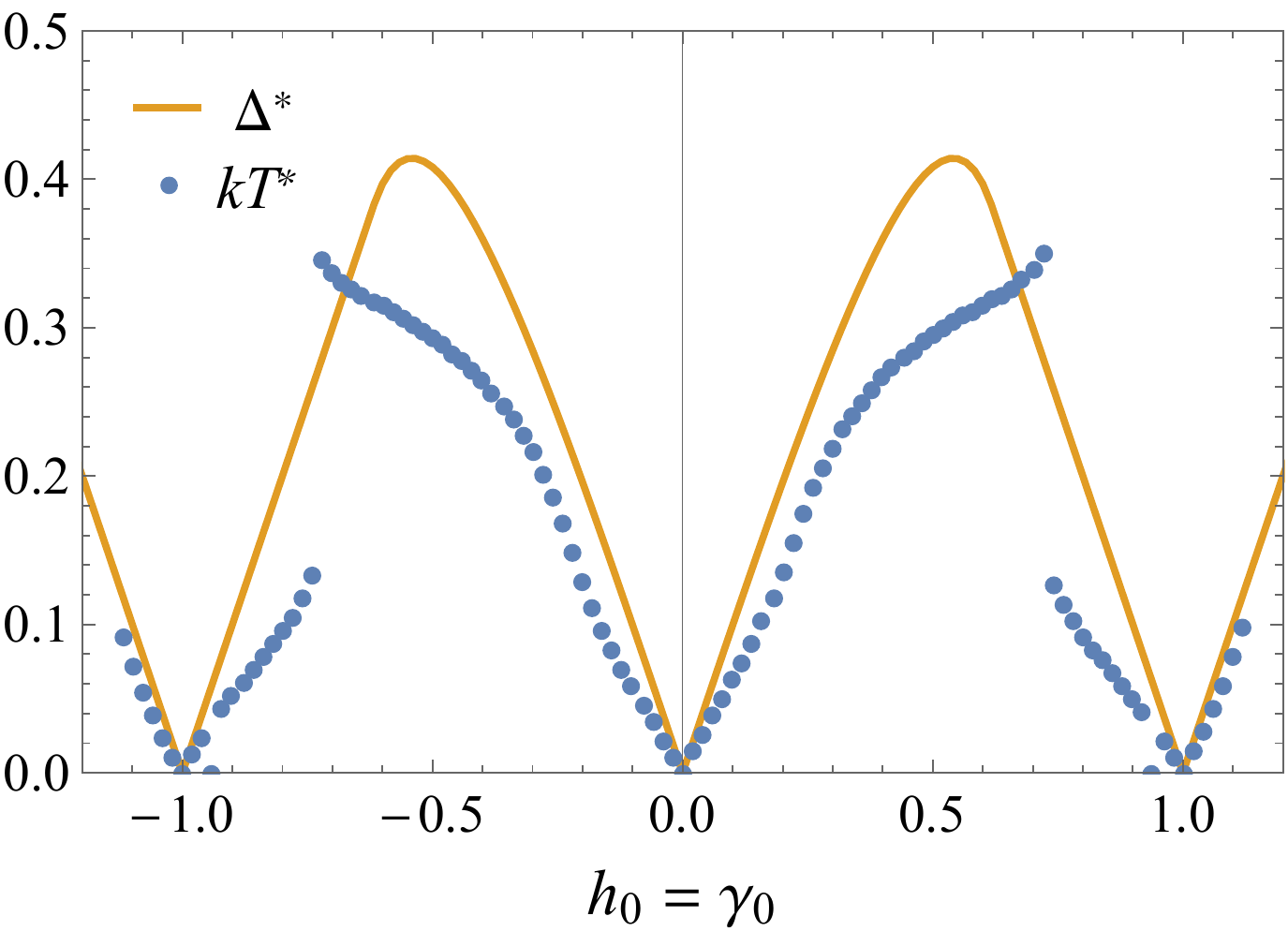}	

\caption{(Color online) In small magnitude quenches $\delta h = \delta \gamma = 0.01$ along the $h_0 = \gamma_0$ line, the irreversible work peaks at some non-zero temperature $T^*$ away from criticality (top). The black arrow indicates increasing $h_0 = \gamma_0$ from 0.0 to 1.5. The excitation gap $\Delta^*$ is of the order of $kT^*$ (bottom).}\label{fig:tstar}
\end{figure}

Additionally, an early reviewer of this paper has pointed out that in Fig.~\ref{fig:irrentropy} the irreversible work per spin $w_\irr$ does not have a monotonic dependence on temperature. Since the irreversible work is the difference of two quantities (the average work done and effective free energy change) with different temperature dependence, it is reasonable to anticipate such non-monotonic behavior. Indeed, observations from several numerical evaluations of the irreversible work reveals that $w_\irr$ peaks at some non-zero temperature $T^*$ away from criticality. For several small magnitude quenches, including the previous case along the $h_0 = \gamma_0$ line (Fig.~\ref{fig:tstar}), we have found that the magnitude of the characteristic thermal energy $kT^*$ is of the order of the energy gap $\Delta^*$. To explain this observation in physical terms, we propose that small thermal fluctuations can lead to increased defect production by supplying the system with some initial energy to overcome the excitation barrier. However, further increases in temperature far above $T^*$ has the opposite effect until $w_\irr$ is completely suppressed at infinite temperature, as expected. We also observe in Fig.~\ref{fig:tstar} that there is a discontinuity in $T^*$ that is caused by two competing peaks in the graph of $w_\irr$ versus temperature. We found it difficult to explain this feature, but our numerical results suggest that this bimodality is due to the separate contributions of the field change $\delta h$ and anisotropy change $\delta \gamma$ to the average work done \eqref{eq:averageworkthermo}. That is, in the examples that we have studied there is no sharp change in $T^*$ as a function of the quench parameters when either of the two terms in Eq.~\eqref{eq:averageworkthermo} are absent or neglected.

\section{Work fluctuations}\label{workfluc}

\begin{table}[bt]
\begin{ruledtabular}
\begin{tabular}{lcc}
Phase of $H_0$ & Poles inside $\mathcal{C}$ & Poles on $\mathcal{C}$\\
\hline
FX, $\sabs{h_0}<1$, $\gamma_0>0$ &  $0$, $z_1$, $z_2$& none\\
FY, $\sabs{h_0}<1$, $\gamma_0<0$ &  $0$, $z_3$, $z_4$& none\\
P$+$, $h_0 >1$   				&  $0$, $z_2$, $z_3$ & none\\
P$-$, $h_0 <-1$   				&  $0$, $z_1$, $z_4$ & none\\
Crit.~$\gamma_0 =0$ line, $\sabs{h_0}<1$& 0 & $z_2=z_3$, $z_1 = z_4$\\
Crit.~field $h_0=1$, $\gamma_0 > 0$ & 0, $z_2$ & $z_1=z_3$\\
Crit.~field $h_0=1$, $\gamma_0 < 0 $ & 0, $z_3$ & $z_2 = z_4$ \\

Crit.~field $h_0= - 1$, $\gamma_0 > 0$ & 0, $z_1$ & $z_2=z_4$\\

Crit.~field $h_0= - 1$, $\gamma_0 < 0$ & 0, $z_4$ & $z_1 = z_3$ \\

Multicritical points & 0 & $z_1=z_2=z_3=z_4$
\end{tabular}
\end{ruledtabular}
\caption{Pole structure of the integrand $f(z)$ for the work fluctuation per spin $\sigma^2$ in a ground state quench \eqref{contourfluctuation}. The pre-quench parameters $h_0$ and $\gamma_0$ of $H_0$ are finite and the post-quench Hamiltonian $H_1$ is not critical.}\label{polestable}
\end{table}

When the initial state is the ground state of $H_0$ the work fluctuation is
\begin{equation}
\Sigma^2 = \frac{1}{4}\sum_{n=0}^{N-1} \epsilon_1(q_n)^2\,\sin^2 2\Delta_n.
\end{equation}
In the thermodynamic limit the fluctuation per spin $\sigma^2 = \lim_{N\to\infty}\Sigma^2/N$ can be represented as a contour integral over a positively oriented unit circle $\mathcal{C}$ via the transformation $e^{ik}\to z$:
\begin{align}
\sigma^2 &= \int_0^{2\pi}\negmedspace \frac{[\gamma_1(h_0-\cos k)-\gamma_0(h_1-\cos k)]^2}{\epsilon_{0}^2(k)}\frac{\sin^2\negthinspace k\,dk}{8\pi},\nonumber\\ 
	&= \frac{1}{2\pi i}\oint_\mathcal{C}f(z)\,dz, \quad (N\to\infty),\label{contourfluctuation}
\end{align}
with integrand 
\begin{equation}
	f(z) =\frac{1}{16}\frac{(z^2-1)^2\prod_\pm (z-z_\pm)^2}{z^3\prod_i(z-z_i)}.
\end{equation}
The zeros $z_\pm$ of the numerator are
\begin{equation}
 z_\pm = \frac{\gamma_0h_1 - \gamma_1h_0 \pm\sqrt{(\gamma_0 h_1-\gamma_1 h_0)^2-(\gamma_0 -\gamma_1)^2 }}{\gamma_0 -\gamma_1},	
\end{equation}
while the (generally) non-zero poles $z_i$ are
\begin{align}
	z_{1} &= \frac{h_0 + \sqrt{\gamma_0^2 +h_0^2 -1}}{\gamma_0+1}, \\
	z_{2} &= \frac{h_0 - \sqrt{\gamma_0^2 +h_0^2 -1}}{\gamma_0+1}, \\
	z_{3} &= \frac{-h_0 + \sqrt{\gamma_0^2 +h_0^2 -1}}{\gamma_0-1},\\
	z_{4} &= \frac{-h_0 - \sqrt{\gamma_0^2 +h_0^2 -1}}{\gamma_0-1}.
\end{align}

\begin{figure}[tb]
\centering
\includegraphics[width=.48\linewidth]{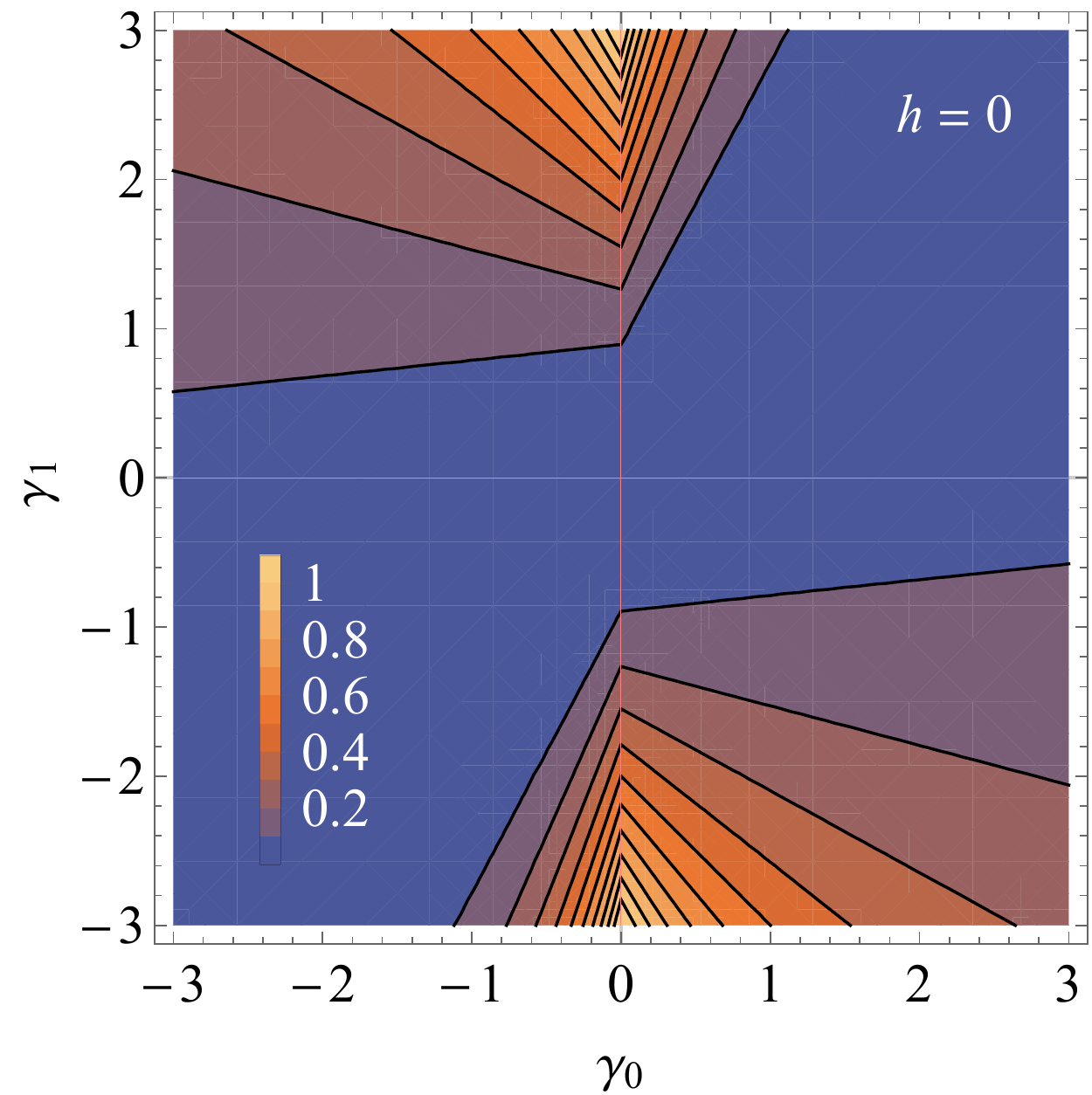}
\includegraphics[width=.48\linewidth]{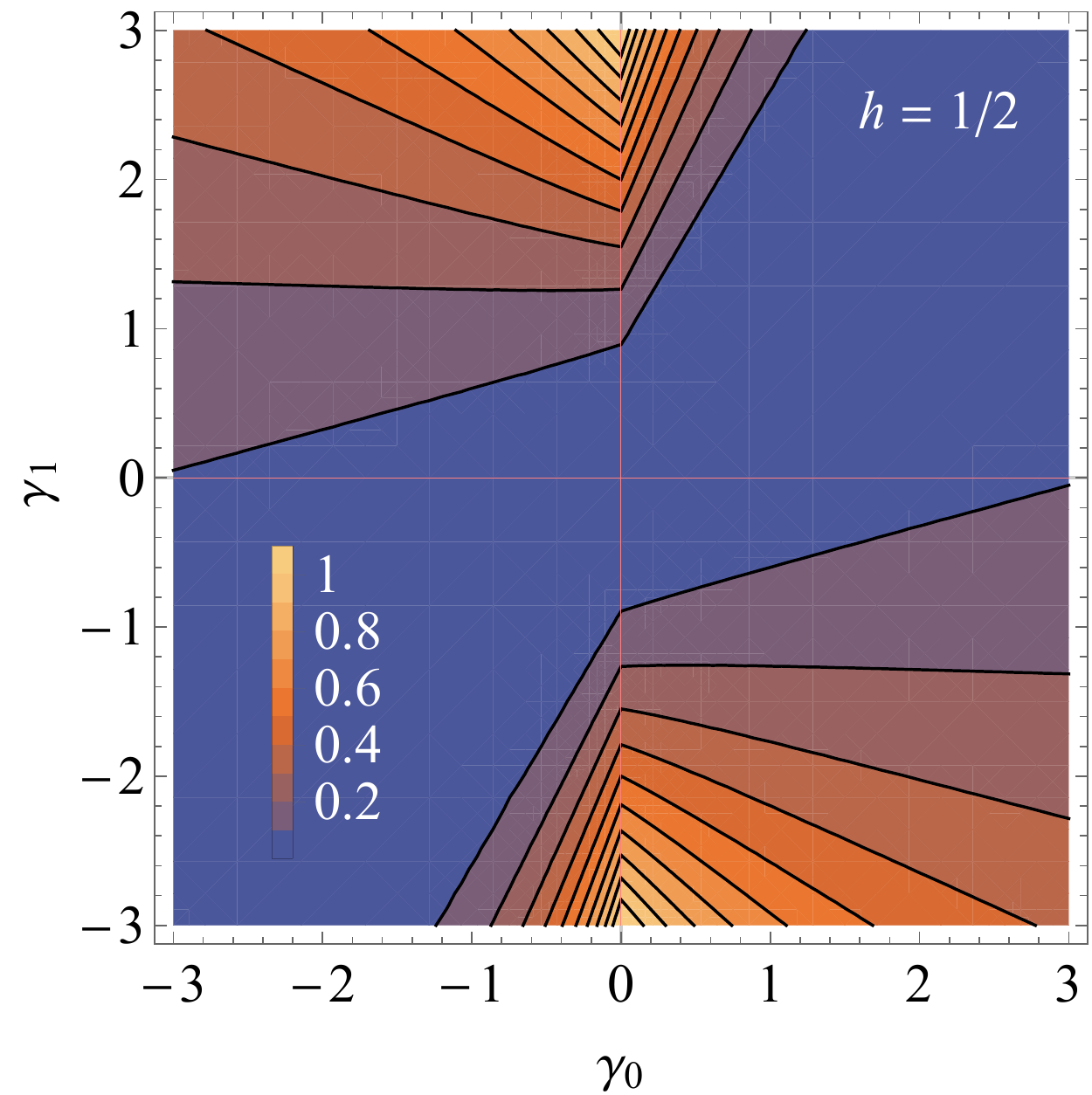}\\
\includegraphics[width=.48\linewidth]{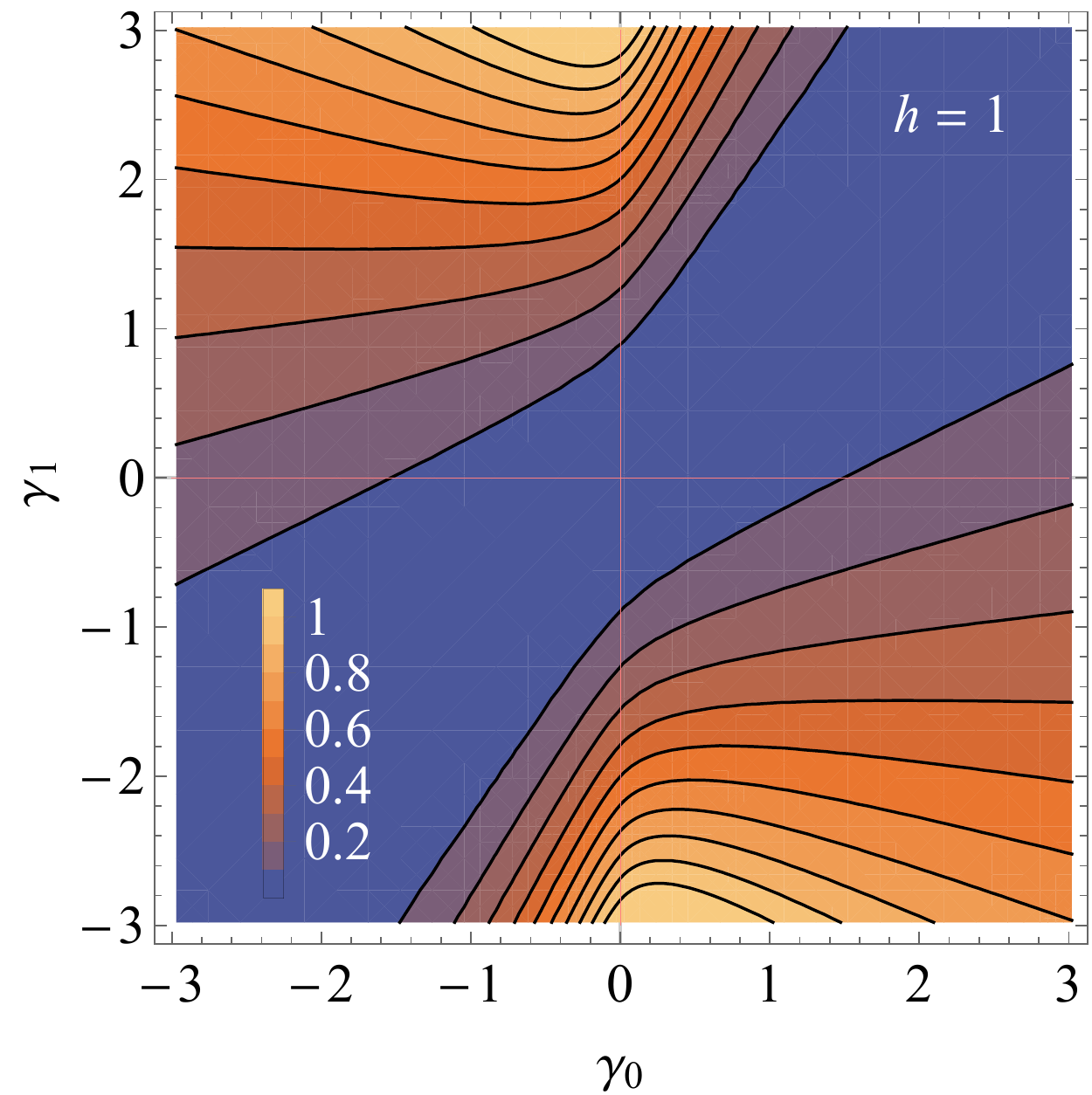}
\includegraphics[width=.48\linewidth]{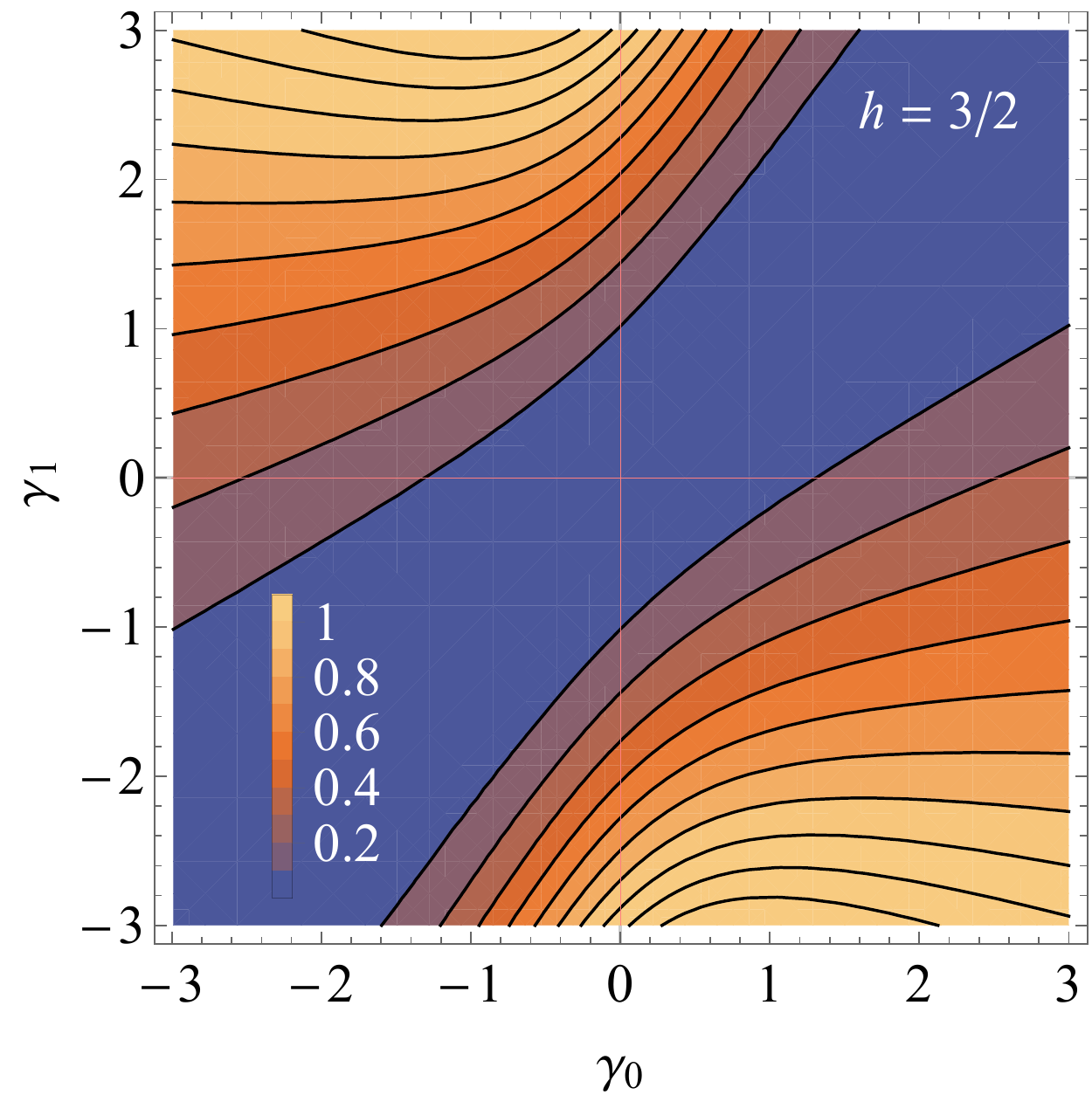}
\caption{(Color online) The work fluctuation per spin $\sigma^2$ for ground state anisotropy quenches at different fixed transverse field ($h=0,\tfrac{1}{2},1,\tfrac{3}{2}$) is not analytic when the pre-quench anisotropy is $\gamma_0 = 0$ and $\abs{h}< 1$. This behavior reflects quantum criticality in the gapless XX region of the XY phase diagram.}  \label{fig:constantfield}
\end{figure}

It turns out that the contour integral \eqref{contourfluctuation} simplifies to the sum of at most three residues of $f$ for non-critical $H_0$:
\begin{equation}\label{eq:exactfluc}
	\sigma^2 = \sum_{\sabs{z_i} < 1}\Res_{z=z_i} f(z).
\end{equation}
This exact analytic result is one of the main contributions of this paper. It allows us to make a general statement about the location of non-analyticities of the work fluctuation $\sigma^2$ on the $(h_0,\gamma_0)$-plane: The work fluctuation is generally not analytic at the points $(h_0,\gamma_0)$ where $H_0$ is critical. An in-depth analysis of the pole structure of the integrand $f(z)$ proves this claim. As summarized in Table \ref{polestable}, a pole moves from inside the contour $\mathcal{C}$ to outside it and another moves from outside the contour to inside it as $H_0$ is changed across a critical point. Since the poles $z_i$ are distinct away from the phase boundaries, the function $\sigma^2(h_0,\gamma_0)$ has different functional forms within each phase of $H_0$. The work fluctuation itself is continuous on the critical lines of $H_0$ (the poles that swap locations become equal at the critical points), but it is not infinitely differentiable along these critical lines. That quantum criticality occurs when the poles $z_i$ fall on the unit circle is reminiscent of the Lee--Yang mechanism of phase transitions in the classical Ising model \cite{lee1952a}, and the analogies between these two phenomena are currently being investigated.

We also make the additional observation that field quenches restricted to the XX line $\gamma_0=\gamma_1 = 0$ have no work fluctuation \eqref{contourfluctuation}. In the isotropic XX regime the transverse field merely acts as a chemical potential for the fermionic quasiparticle excitations of the model. Thus, energy eigenstates of an XX Hamiltonian with field $h_0$ are also eigenstates of any post-quench XX Hamiltonian with a different field $h_1$. The resulting probability distribution for the work done is therefore delta-peaked at $W = \Delta E = -\tfrac{1}{2}N(h_1-h_0)$ with no work fluctuation for any magnitude of field quench. 

\textit{Anisotropy quenches---}With an exact formula for the work fluctuation at hand, we can consider specific examples where the presence of non-analyticities reveal the criticality of the pre-quench Hamiltonian $H_0$. For instance, we take the case of anisotropy quenches $\gamma_0\to\gamma_1$ at different fixed fields $h_0=h_1 = h$. As shown in Fig.~\ref{fig:constantfield} the work fluctuation has a kink (finite discontinuity in the first derivative with respect to $\gamma_0$) at the isotropic line $\gamma_0 = 0$ only for field magnitudes below the critical threshold $\abs{h_0}<1$. An explicit calculation for the case of quenches toward the Ising line ($\gamma_1 = 1$) gives
\begin{equation}
	\frac{\partial \sigma^2}{\partial \gamma_0}\biggr|_{\gamma_0 \to 0^+} - \frac{\partial \sigma^2}{\partial \gamma_0}\biggr|_{\gamma_0 \to 0^-} = \begin{cases}
		(h^2 - 1)/2, & \sabs{h}<1,\\
		0,& \sabs{h}>1,
	\end{cases}
\end{equation}
and we find that the kink only appears along the critical XX line.

\textit{Ground state quench along the Ising line---}Another example in which our solution yields concrete results is given by a ground state quench of the transverse field Ising model. The work fluctuation per spin has a discontinuous field derivative at pre-quench critical fields $h_0 = \pm 1$ (except when $h_1 = \pm 1$) \cite{bayocboc2015a}:
\begin{equation}\label{isingfluc}
	\sigma^2 = \begin{cases}
	(h_1-h_0)^2/8, & \sabs{h_0}<1,\\
	(h_1-h_0)^2/(8\sabs{h_0}), & \sabs{h_0}\ge 1.
	\end{cases}
\end{equation}
In the exceptional case $h_1 = \pm 1$ the non-analyticity at criticality is more subtle. A series expansion of $\sigma^2$ about $h_0 = \pm 1 = h_1$ reveals that it is the third derivative $\partial^3(\sigma^2)/\partial h_0^3$ that first becomes discontinuous at $h_0 = \pm 1$. The exact result \eqref{isingfluc} includes the previously calculated approximate asymptotic formula obtained for $h_1 \gg 1$ \cite{silva2008a}.

\begin{figure}[tb]
\centering
\subfloat[Quench to $(h_1,\gamma_1) = (0,1)$.\label{sfig:zerofieldTIM}]{%
  \includegraphics[width=.48\linewidth]{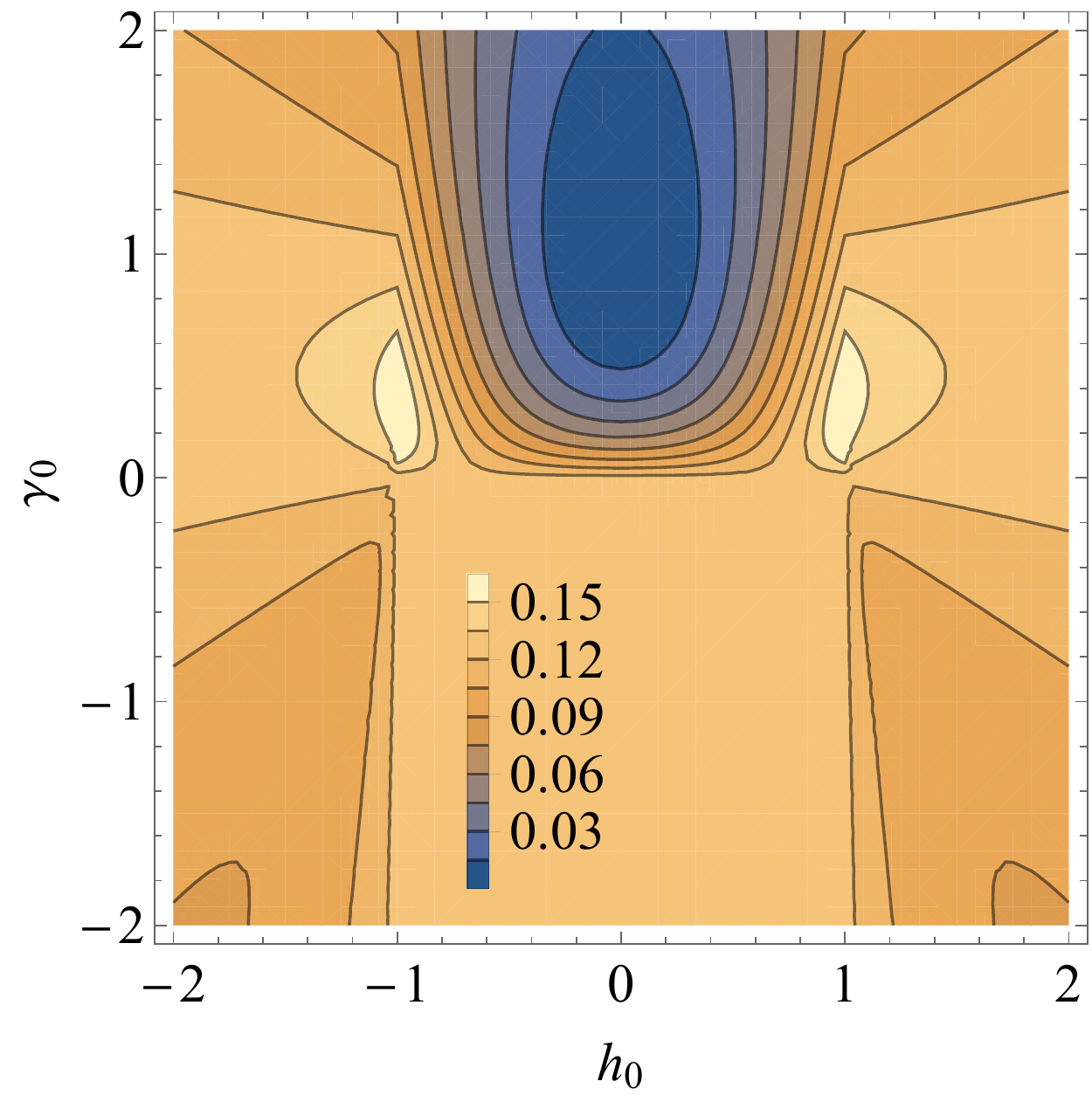}%
}\hfill
\subfloat[Quench to $(h_1,\gamma_1) = (1,0)$.\label{sfig:multicrit}]{%
  \includegraphics[width=.48\linewidth]{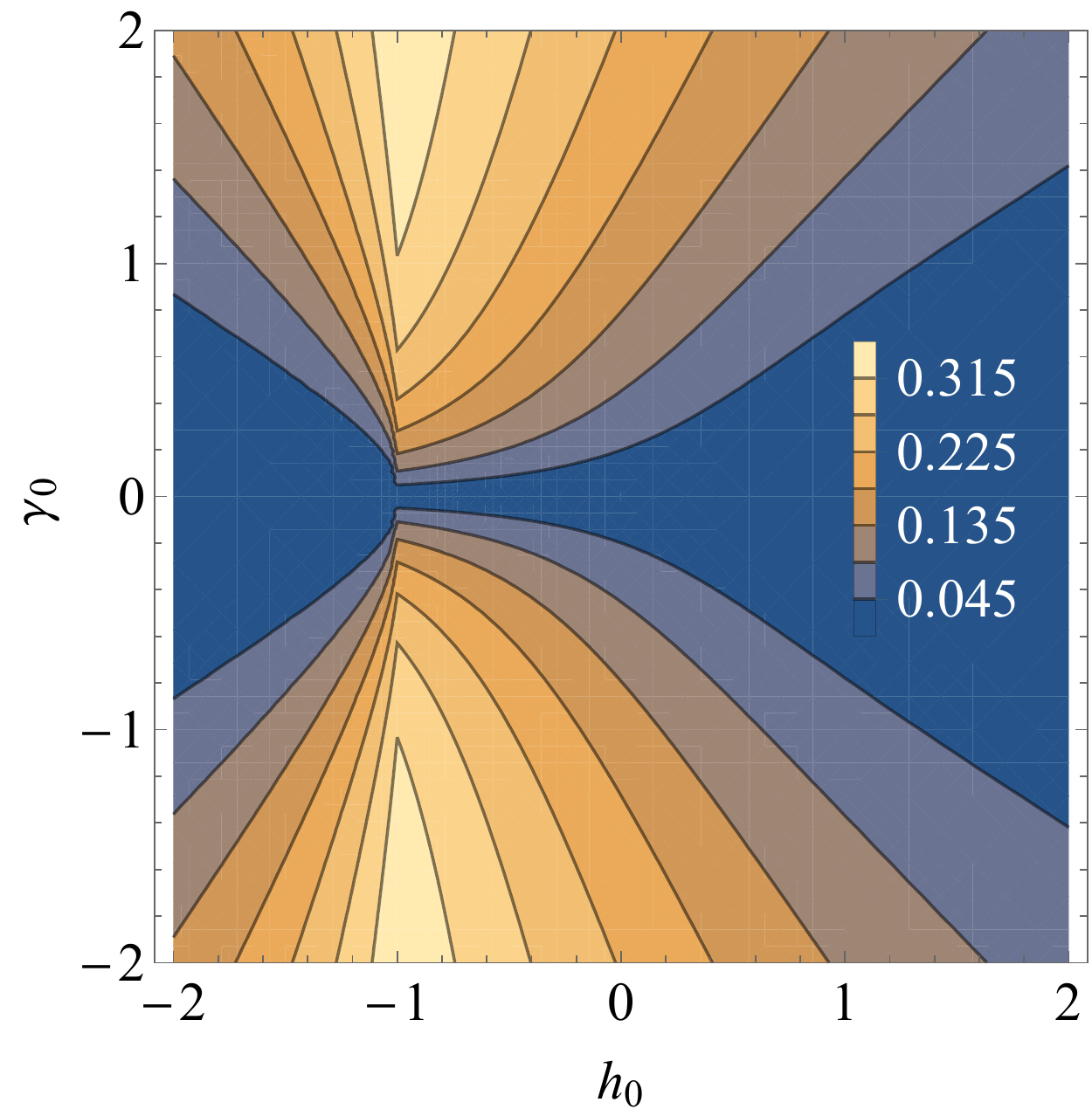}%
}
\caption{(Color online) Work fluctuation per spin in the thermodynamic limit $\sigma^2 = \lim_{N\to \infty}\Sigma^2/N$ for ground state quenches that end at (a) the zero-field Ising model, and (b) the multicritical point at $h_1 = 1$. The work fluctuation is not analytic along the critical field lines $\sabs{h_0}=1$ and the critical XX line ${\gamma_0}=0, \sabs{h_0}<1$.}\label{fig:workfluc_final}
\end{figure}

\textit{Quench with fixed post-quench parameters---}The non-analytic behavior of the work fluctuation when the pre-quench Hamiltonian is critical is further illustrated in examples where the post-quench parameters $(h_1,\gamma_1)$ are fixed. In Fig.~\ref{sfig:zerofieldTIM} we show our calculations for $\sigma^2$ in the case of a ground state quench toward the zero-field transverse Ising model, which clearly shows kinks in the work fluctuation along the critical field and critical isotropic regions. The exact analytic solution used to obtain this plot is given in Appendix~\ref{app:zeroTIM}. 

We now discuss in some detail a more involved example in which the post-quench Hamiltonian $H_1$ lies on the multicritical point $(h_1 ,\gamma_1)= (1,0)$, so that both $H_0$ and $H_1$ are quantum critical. The work fluctuation in this case is graphed in Fig.~\ref{sfig:multicrit}, which reveals kinks along the pre-quench critical field $h_0 = -1$. The difference in the field derivatives of the work fluctuation per spin along this critical line is
\begin{equation}
 \frac{\partial \sigma^2}{\partial h_0}\biggr|_{h_0\to -1^+} - \frac{\partial \sigma^2}{\partial h_0}\biggr|_{h_0\to -1^-} = -\frac{1}{\sabs{\gamma_0}}.
 \end{equation} 
The first derivative of the work fluctuation with respect to anisotropy is also discontinuous across the critical XX line:
\begin{equation}
 \frac{\partial \sigma^2}{\partial \gamma_0}\biggr|_{\gamma_0\to 0^+} - \frac{\partial \sigma^2}{\partial \gamma_0}\biggr|_{\gamma_0\to 0^-} = \begin{cases}
	(1-h_0)^2/2, & \sabs{h_0}\le 1,\\
	 0, & \sabs{h_0}> 1,
 \end{cases}.
 \end{equation} 

However, the non-analytic behavior of $\sigma^2$ at $h_0 = 1$ is more subtle. Inspection of Fig.~\ref{sfig:multicrit} does not immediately reveal singular derivatives at $h_0 = 1$. In Appendix~\ref{app:multi} we prove that the first discontinuous derivative of $\sigma^2$ with respect to the field at $h_0 = 1$ is, in fact, the fifth. That non-analyticities in $\sigma^2$ are weakened when $h_0 = h_1$ fall on the same critical region can be traced back to the contour integral formula \eqref{contourfluctuation} and Table \ref{polestable}. We find that the pole that contributes to the non-analytic behavior at the critical field $h_0$ is cancelled by one of the terms $(z - z_\pm)^2$ in the numerator of the integrand $f(z)$ when $h_1=h_0$. 

\begin{figure}[tb]
\centering
  \includegraphics[width=.7\linewidth]{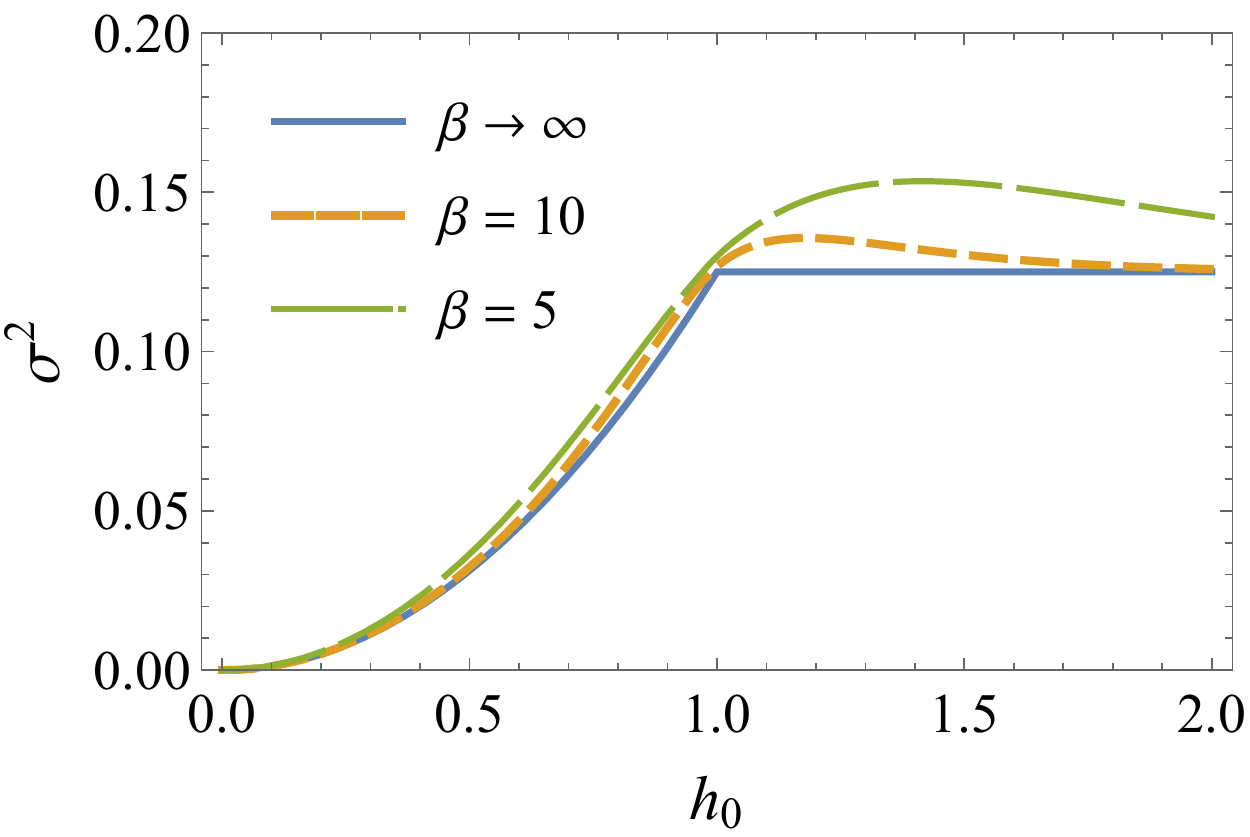}

  \includegraphics[width=.7\linewidth]{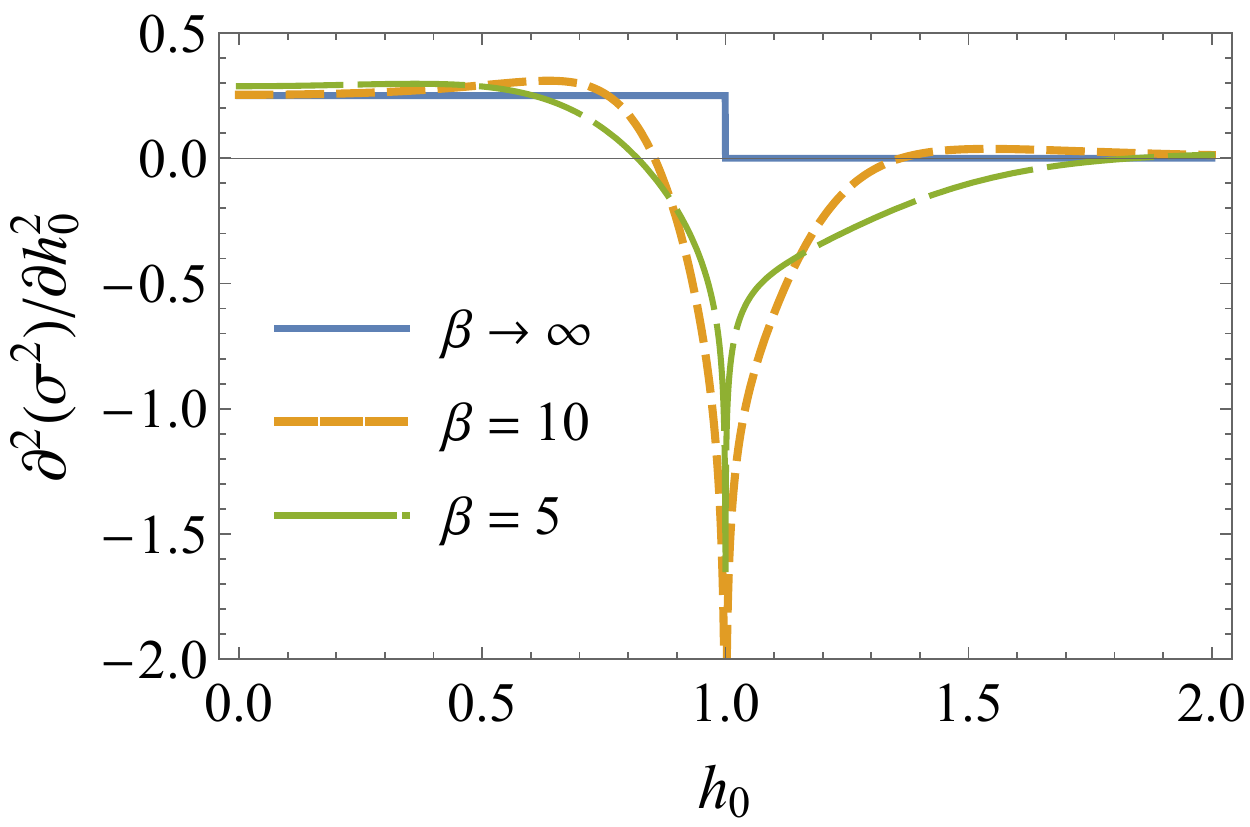}%

\caption{(Color online) For quenches along the Ising line $\gamma_0 = \gamma_1 = 1$ toward zero-field $h_1 = 0$, an initial non-zero temperature smoothens the kink of the work fluctuation per spin at the critical pre-quench field $h_0 = 1$ (top). However, the numerically calculated second derivative of $\sigma^2$ at non-zero temperature suggests the existence of a cusp singularity at $h_0 = \pm 1$ (bottom).}\label{fig:fluctemp}
\end{figure}

\textit{Non-zero temperature---}Extending these results to initially mixed states at finite $\beta$ yields:
\begin{multline}
\sigma^{2} = \int_{0}^{2\pi} \biggl\{\frac{(h_0-h_1)^2 + (\gamma_0 - \gamma_1)^2\sin^2 k}{2\cosh\bigl[ {\beta\epsilon_{0}(k)}/{2} \bigr]} \\
 \quad +   \biggl(\frac{[\gamma_1 h_0-\gamma_0 h_1 + (\gamma_0-\gamma_1)\cos k ]^2\sin^2 k}{\epsilon_{0}^2(k)}  \\ 
\times\tanh \bigl[ {\beta\epsilon_{0}(k)}/{2}\bigr]  \biggr) \biggr\}  \frac{dk}{8\pi}  .
\end{multline}
We see that singular behavior that can potentially arise from the criticality of $H_0$ must come from the second term enclosed in large parentheses, which has $\epsilon_0^2$ in the denominator. Near criticality, the temperature-dependent factor $\tanh ({\beta\epsilon_{0}}/{2})$ only partially cancels the divergence caused by vanishing $\epsilon_0^2$. We have performed some numerical calculations of this term and found that $\sigma^2$ is generally smooth (with continuous first derivatives) at critical $H_0$. However, as in the cases of weakened non-analyticities above, we have found numerical evidence that higher derivatives of $\sigma^2$ are singular at the critical points even at non-zero temperatures. In Figure \ref{fig:fluctemp} we present the case of a non-zero temperature quench along the Ising line with post-quench field $h_1 = 0$. The fluctuation per spin is seen to be continuous at $h_0 = 1$, while its second derivative with respect to $h_0$ diverges. This result is noteworthy as the singular behavior of observables at quantum critical points tend to be washed out by thermal fluctuations.

\section{Concluding remarks}
In this paper, we have studied the statistics of the  work done in quantum quenches of the one-dimensional XY model. We obtained the characteristic function of the probability distribution of work for a generic quench from initial values $h_0$ and $\gamma_0$ to final values $h_1$ and $\gamma_1$ of the magnetic field and anisotropy, respectively. The average work done during the quench was calculated from this characteristic function and sharp peaks in irreversible entropy production were observed in the vicinity of the quantum critical lines. In small magnitude field quenches, irreversible entropy production was shown to be enhanced (reduced) by increased anisotropy in the paramagnetic (ferromagnetic) phase.  We also derived an exact contour integral formula for the work fluctuation per spin in ground state quenches and showed how non-analyticities can arise at points where the pre-quench XY Hamiltonian is critical. Finally, we obtained exact and compact solutions for the work statistics in the case of ground state quenches along the the transverse field Ising line. This latter result allowed us to identify logarithmic singularities and finite discontinuities in the field derivatives of the average work and work fluctuation, respectively, at the quantum critical pre-quench fields.

\textit{Note added---}Shortly after the submission of this manuscript a related paper was published reporting similar numerical results for finite-length quenched XY chains in the presence of three-site interactions \cite{zhong2015a}.

\acknowledgments
The authors are supported by the University of the Philippines OVPAA through Grant No.~OVPAA-BPhD-2012-05, and the University of the Philippines Diliman OC and OVCRD through Project No.~141420 PhDIA. FB Jr.~acknowledges support from the DOST Science Education Institute through its ASTHRD Program.

\appendix
\section{Quench to zero-field Ising model}\label{app:zeroTIM}
When the post-quench parameters are $(h_1,\gamma_1) = (0,1)$, which correspond to the zero-field transverse field Ising model, the residue formula \eqref{eq:exactfluc} gives
\begin{equation}
	\sigma^2 = \begin{cases}
	\dfrac{(1 - \gamma_0)^2}{8(1 + \gamma_0)^2} + \dfrac{\gamma_0 h_0^2}{(1 + \gamma_0)^3}, & \text{FX},\\
	1/8, & \text{FY},\\
	\dfrac{1 + \gamma_0(1+4h_0^2) + \gamma_0^2 + \gamma_0^3}{8(1+\gamma_0)^3} & \\
	\quad \pm\dfrac{2\gamma_0 h_0 [2(1-h_0^2) +\gamma_0 -\gamma_0^2]}{8(1+\gamma_0)^3\sqrt{\gamma_0^2 +h_0^2-1}}, & \text{P$\pm$}.
	\end{cases}\label{eq:zeroTIM}
\end{equation}
In the paramagnetic region $\sigma^2$ has a Taylor series expansion about $\gamma_0 = -1$ and is therefore infinitely differentiable with respect to $\gamma_0$, that is, $\gamma_0=-1$ is a removable singular point.

\section{Quench to multicritical point}\label{app:multi}
Let us consider ground state quenches toward the multicritical point $(h_1,\gamma_1) = (1,0)$. We use the residue formula \eqref{eq:exactfluc} and expand the work fluctuation about the critical field $h_0 = 1$ in the ferromagnetic and P$+$ paramagnetic phases of $H_0$:
\begin{widetext}
\begin{equation}
	\sigma^2 = \frac{\gamma_0^2}{8(1+\sabs{\gamma_0})^3}\begin{cases}
	1 + 3 \sabs{\gamma_0 } - 4(h_0 - 1) +  \dfrac{2(h_0 - 1)^2}{\sabs{\gamma_0}}, & \text{FX and FY,}\\
	1 + 3 \sabs{\gamma_0 } - 4(h_0 - 1) +  \dfrac{2(h_0 - 1)^2}{\sabs{\gamma_0}} -  \dfrac{(1+\sabs{\gamma_0})^3(h_0 - 1)^5}{2\sabs{\gamma_0}^7} +\dotsb, & \text{P+}.
	\end{cases}
\end{equation}\label{multififth}
\end{widetext}
We conclude that the work fluctuation and its first four derivatives with respect to $h_0$ are continuous at the critical field $h_0 = 1$. Furthermore, at this critical field the fifth derivative has the discontinuity
\begin{equation}
	\frac{\partial^5 (\sigma^2)}{\partial h_0^5}\biggr|_{h_0 = 1^+} - \frac{\partial^5 (\sigma^2)}{\partial h_0^5}\biggr|_{h_0 = 1^-} = -\frac{15}{2 \sabs{\gamma_0}^5}.
\end{equation}

%\bibliography{xyquenchbib}


\end{document}